\documentstyle[12pt]{article}

\begin{document}
\thispagestyle{empty}

\begin{center}
\LARGE \tt \bf{Stretching Riemannian spherical solar dynamos from
differential rotation}
\end{center}

\vspace{1.5cm}

\begin{center} {\large L.C. Garcia de Andrade\footnote{Departamento de
F\'{\i}sica Te\'{o}rica - Instituto de F\'{\i}sica - UERJ

Rua S\~{a}o Fco. Xavier 524, Rio de Janeiro, RJ

Maracan\~{a}, CEP:20550-003 , Brasil.E-mail:garcia@dft.if.uerj.br.}}
\end{center}

\vspace{2.0cm}

\begin{abstract}
Stretching solar dynamos from differential rotation in a Riemannian
manifold setting is presented. The spherical model follows closely a
twisted magnetic flux tube Riemannian geometrical model or flux rope
in solar physics, presented previously by Ricca [Solar Physics
(1997)]. The spherical model presented here present new and
interesting feature concerning its connection with spherical steady
solar dynamos. One of this new feature is represented by the fact
that the by considering poloidal magnetic field component much
weaker than its toroidal counterpart, one obtains a stretch dynamo
action where the Riemannian solar spherical line element is
proportional to differential rotation. This result is obtained also
by using the Vainshtein-Zeldovich stretch, twist and fold (STF)
method to generate dynamos. One notes that for high magnetic
Reynolds of $R{m}=O({10^{7}})$ the dynamo action is present for a
corresponding small stretching factor of $K^{2}=1.6$ where $K^{2}=1$
represents the unstretched dynamo. The constant stretched dynamos
considered here are shown to be Riemann-flat, where the Riemann
curvature tensor vanishes. Solar cycle dynamos are therefore
compatible with the Riemann-flat stretching dynamo model from solar
differential rotation presented here.
\end{abstract}
\newpage
\section{Introduction}
Stretching in convective overshooting regions in solar physics has
been previously investigated by Fisher at al \cite{1} by stretching
magnetic flux tubes \cite{2} from the differential rotation of the
Sun. The differential rotation acts to strengthen magnetic fields
with a non-azimuthal component by strengthing them in the azimuthal
direction. It is a consensus that dynamos responsible for generating
solar cycles reside in the form of flux tubes or flux tube dynamos
\cite{1}. Have been convinced that dynamos can reside in the for of
such Riemannian structures as thin magnetic twisted flux tubes,
Ricca \cite{3} developed a Riemannian geometrical model of twisted
magnetic flux tubes to explain inflexional and desiquilibrium
properties of these magnetic tubes. Other models of Riemannian
curvature in material flow lines have been also displayed recently
\cite{4}. In this paper following the importance of spherical solar
models \cite{5} it is shown that it is possible to obtain a
analogous Riemannian geometrical \cite{6} model in spherical
coordinates where, analytical solutions are obtained of the
self-induction steady equations. The first result to be shown is
that by using the Vainshtein-Zeldovich \cite{7} stretch-twist-fold
\cite{8} generation method it is possible to obtain a ratio between
the Riemannian lines elements of the stretched and unstretched
spherical models and connect it to the differential rotation.
Actually on a recent paper we call the attention to the fact that is
possible to obtain a analytical Riemannian geometrical model of
conformal stretching dynamo in highly conductively Riemannian
manifold flow. It is also shown here that unstretched spherical
Riemann curvature tensor vanishes, which mathematicians call a
Riemann-flat manifold, which leads to a nondynamo or by the absence
of dynamo action. Actually finally we show that our model is
compatible with a solution given by Livermore and al \cite{9} of the
role of stretching in spherical shells, since our dynamo possesses a
high order magnetic Reynolds number $Rm=O(10^{7})$, whereas dynamo
action considered by them has a Reynolds number of maximum
$Rm=O(500)$. Other dynamo mechanism has been recently displayed in
conformal stretching \cite{10} in Riemannian manifolds. This paper
is organized as follows: In section II the model is presented and
the Riemannian curvature tensor is computed for the general case of
non-uniform stretching. In section III the Riemann-flat uniform
stretching model is addressed along with the self-induction equation
for the steady dynamo solar flow also with solar physics
implications and discussions are presented in section IV.
\section{Riemannian non-uniformly stretching}
This section presents the Riemann metric of a spherical symmetric
space in curvilinear coordinates $(r,{\theta},{\phi})$. This metric
is encoded into the Riemann line element
\begin{equation}
ds^{2}=
dr^{2}+r^{2}[d{{\theta}}^{2}+K^{2}sin^{2}{\theta}d{\phi}^{2}]
\label{1}
\end{equation}
where, accordingly to our hypothesis , K is a constant uniform
stretch. As one can see bellow, the differential rotation \cite{11}
is constant along the toroidal direction and slowly decays along the
azimuthal poloidal direction. Just by analogy the Riemann metric of
the magnetic flux tube \cite{3} is appended
\begin{equation}
d{l}^{2}= dr^{2}+r^{2}d{{\theta}_{R}}^{2}+K^{2}(r,s)ds^{2} \label{2}
\end{equation}
Note that these two Riemann metrics are similar, however while
Ricca's metric helps us to investigate the topology of solar loops,
the present metric helps us to investigate differential rotation of
sun and other higher density stars. Let us now consider the
magnetohydrodynamic (MHD) equations
\begin{equation}
{\nabla}.\textbf{B}=0\label{3}
\end{equation}
which also has its solenoidal counterpart in the incompressibility
of dynamo flow case, as
\begin{equation}
{\nabla}.\textbf{v}=0\label{4}
\end{equation}
and self induction equation
\begin{equation}
{\partial}_{t}\textbf{B}+{\nabla}{\times}(\textbf{v}{\times}\textbf{B})={\eta}{\nabla}^{2}\textbf{B}\label{5}
\end{equation}
Here we consider the non-radial fields as
$\textbf{B}=\textbf{e}_{\phi}{B}_{\phi}(r,{\theta})+\textbf{e}_{\theta}B_{\theta}(r,{\theta})$
which is the sum of the toroidal and poloidal magnetic field
components respectively. Note that since the solar flow considered
is steady, and highly conductive, or null magnetic diffusivity,
$({\eta}=0)$, this equation reduces to
\begin{equation}
{\partial}_{t}\textbf{B}+{\nabla}{\times}(\textbf{v}{\times}\textbf{B})=0\label{6}
\end{equation}
Since the dynamo is also considered to be steady, the dynamo
equation reduces to
\begin{equation}
{\nabla}{\times}(\textbf{v}{\times}\textbf{B})=0\label{7}
\end{equation}
A so-called "poor" solutions of this simple equation can be obtained
by considering solutions such as
\begin{equation}
{\textbf{v}}{\times}\textbf{B}={\nabla}{\phi}\label{8}
\end{equation}
where ${\phi}$ is a scalar component. A particular case of this is
given by the case where $\textbf{v}$ and $\textbf{B}$ are parallel.
These cases are of no interest for us here and instead we shall
address the case of expanding the LHS of equation (\ref{7}) as
\begin{equation}
({\nabla}.\textbf{v})\textbf{B}=({\nabla}.\textbf{B})\textbf{v}\label{9}
\end{equation}
and further develope this equation in the background of stretched
curved Riemannian three-dimensional spherical space given by Riemann
metric (\ref{1}). Before we dig into these equations and solve them
let us consider here the Riemannian curvature of two general cases ,
namely when $K^{2}$ stretching factor depends upon either on
$coordinate-{\theta}$ or on $coordinate-{\phi}$. In the second case
as is easily computed with the tensor package of general
relativistic package simply adapted for the three-dimensional
Riemannian space, in any computer software, one obtains that the
Riemann tensor vanishes, while in the first case the Riemann
curvature reads
\begin{equation}
\textbf{R}_{{\theta}{\phi}{\theta}{\phi}}=-\frac{r^{2}}{2}[sin{2{\theta}}\frac{dK^{2}}{d{\theta}}+sin^{2}{\theta}\frac{d^{2}K^{2}}{d{\theta}^{2}}
-\frac{1}{2K^{2}}(\frac{dK^{2}}{d{\theta}})^{2}sin^{2}{\theta}]\label{10}
\end{equation}
note that in the uniform stretching case, where $K^{2}=constant$
this curvature vanishes what justifies to call such a solar dynamo
manifold in spherical coordinates, Riemann-flat. Actually from
expression (\ref{2}),and since Ricca´s stretching factor
$K^{2}=K^{2}(r,s)$ is non-uniform it is easy to show that the
Riemann curvature tensor components in the case of twisted magnetic
flux tube, with strong foldingv (curvature) ,as in dynamo theory is
\begin{equation}
\textbf{R}_{1313}=\textbf{R}_{rsrs}=
-\frac{1}{4K^{2}}[2K^{2}{\partial}_{r}A(r,s)-A^{2}]=-\frac{1}{2}\frac{K^{4}}{r^{2}}=-\frac{1}{2}r^{2}{\kappa}^{4}cos^{2}{\theta}
\label{11}
\end{equation}
and
\begin{equation} \textbf{R}_{2323}=\textbf{R}_{{\theta}s{\theta}s}=
-\frac{r}{2}A(r,s)= -{K^{2}}\label{12}
\end{equation}
which confirms that the stretch of twisted flux tube dynamos are
really non-uniform as in Fisher at al cases.
\section{The spherical steady solar dynamo}
Let us first write the expression for solenoidal vector $\textbf{B}$
in this Riemannian spherical metric stretched background as
\begin{equation}
{\nabla}.\textbf{B}=\frac{1}{Kr^{2}sin{\theta}}[{\partial}_{\theta}(B_{\theta}Krsin{\theta})+{\partial}_{\phi}(B_{\phi}r)]
=0\label{13}
\end{equation}
which reduces to
\begin{equation}
{\partial}_{\theta}(B_{\theta}sin{\theta})+{\partial}_{\phi}(\frac{1}{K}B_{\phi})=0\label{14}
\end{equation}
by  considering that $B_{\phi}$ toroidal component does not depend
on coordinate ${\phi}$ ${\partial}_{\phi}B_{\phi}$ vanishes and the
absence of last term reduces equation for the poloidal component to
\begin{equation}
{\partial}_{\theta}(B_{\theta}sin{\theta})=0\label{15}
\end{equation}
which yields the immediate solution
\begin{equation}
B_{\theta}=B_{0}(r)csc{\theta}\label{16}
\end{equation}
which is a periodic solution. The equations above and bellow are
computed by using the Riemannian line element in the form
\begin{equation}
ds^{2}=(h_{i}dx^{i})^{2}\label{17}
\end{equation}
where the metric coefficients are
\begin{equation}
h_{1}=1\label{18}
\end{equation}
\begin{equation}
h_{2}=r\label{19}
\end{equation}
\begin{equation}
h_{3}=rKsin{\theta}\label{20}
\end{equation}
Note also that the vorticities of differential rotation obeys the
following equation
\begin{equation}
{\nabla}.\vec{\omega}=0\label{21}
\end{equation}
which yields the following constraints on differential rotation
\begin{equation}
K^{-1}{\partial}_{\phi}{\omega}_{\phi}+sin{\theta}{\partial}_{\theta}{\omega}_{\theta}=0\label{22}
\end{equation}
Assuming that component ${\omega}_{\theta}$ does not depend on
coordinate ${\phi}$ one obtains
\begin{equation}
{\omega}_{\phi}=-K{\phi}sin{\theta}{\partial}_{\theta}ln{\omega}_{\theta}\label{23}
\end{equation}
From the self-induction equation a long but straightforward
computation only tells us that the relation between toroidal and
poloidal components is given by
\begin{equation}
\frac{{B}_{\phi}}{B_{\theta}}=\frac{v_{\phi}}{v_{\theta}}=\frac{{\omega}_{\phi}r}{{\omega}_{\theta}r}\label{24}
\end{equation}
which implies
\begin{equation}
\frac{{B}_{\phi}}{B_{\theta}}=\frac{{\omega}_{\phi}}{{\omega}_{\theta}}\label{25}
\end{equation}
From the Vainshtein et al \cite{13} fractal Riemannian geometry and
stretching
\begin{equation}
\frac{{{B}_{\phi}}^{2}}{{B_{\theta}}^{2}}=\frac{ds^{2}}{{ds_{0}}^{2}}\label{26}
\end{equation}
a similar expression to this one was obtained by Fisher et al in the
case of flux tubes which is given by
\begin{equation}
\frac{{{B}_{\phi}}^{2}}{{B_{\theta}}^{2}}=\frac{l^{2}}{{l_{0}}^{2}}\label{27}
\end{equation}
where l is the length of magnetic flux tube. From the Riemann
metrics of stretched and unstretched solar models one yields
\begin{equation}
\frac{ds^{2}}{{ds_{0}}^{2}}=1+(K^{2}-1)r^{2}sin^{2}{\theta}\frac{{d{\phi}}^{2}}{{ds^{2}}_{0}}\label{28}
\end{equation}
Taking into account the expression for the differential rotation as
${\omega}=\frac{d{\phi}}{dt}$ and $v_{0}=\frac{ds_{0}}{dt}$ which
yields
\begin{equation}
\frac{ds^{2}}{{ds_{0}}^{2}}=1+(K^{2}-1)r^{2}sin^{2}{\theta}\frac{{{\omega}_{\phi}}^{2}}{{{v}_{0}}^{2}}\label{29}
\end{equation}
Vainshtein et al \cite{13} also considered the relation between the
ratio of squared magnetic fields and the magnetic Reynolds number as
\begin{equation}
\frac{{{B}_{\phi}}^{2}}{{B_{\theta}}^{2}}={(R{m})}^{n}\label{30}
\end{equation}
Comparing this equation with expression (\ref{27}) yields
\begin{equation}
\frac{{{B}_{\phi}}^{2}}{{B_{\theta}}^{2}}=1+(K^{2}-1)r^{2}sin^{2}{\theta}\frac{{{\omega}_{\phi}}^{2}}{{v^{2}}_{0}}\label{31}
\end{equation}
From expressions (\ref{29}) and (\ref{30}) one obtains $n=0$
corresponds to $K^{2}=1$ or the unstretched Riemannian metric. This
corresponds to the nondynamo action since the toroidal does not
amplify w.r.t poloidal component. For $n=1$ the relation (\ref{30})
can be computed at the surface of the Sun, where
$R_{sun}\approx{10^{10}} cm$ and the Reynolds number can be
considered as high as $10^{10}$.  In this case the relation between
the two components of the magnetic field is given by
\begin{equation}
\frac{{{B}_{\phi}}^{2}}{{B_{\theta}}^{2}}=1+(K^{2}-1){R_{sun}}\frac{{{\omega}_{\phi}}^{2}}{{{\omega}^{2}}_{0}}\label{32}
\end{equation}
where one has compute it at the ${\theta}=\frac{\pi}{2}$, and thus
\begin{equation}
\frac{{{B}_{\phi}}^{2}}{{B_{\theta}}^{2}}=(K^{2}-1){10}^{10}\frac{{{\omega}_{\phi}}^{2}}{{v^{2}}_{0}}\label{33}
\end{equation}
in the strong stretch limit where ${K^{2}}>>1$. Taking into account
that in the Sun the differential rotation is about $26$ per cent
\cite{13} of solar rotation one obtains
\begin{equation}
\frac{{{\omega}_{\phi}}^{2}}{{{\omega}^{2}}_{0}}\approx{10^{-2
}}\label{34}
\end{equation}
\begin{equation}
\frac{{{B}_{\phi}}}{{B_{\theta}}}=K{\times}{10}^{4}\label{35}
\end{equation}
This shows that , since $K^{2}>1$ to obtain a stretch that induces
an expressive amplification in the toroidal component with the
dynamo action one must have a bound value of the for this expression
of at least
\begin{equation}
\frac{{B}_{\phi}}{B_{\theta}}\ge{{10}^{4}}\label{36}
\end{equation}
This expression is well within observational data since for example
a poloidal field of $0.5 G$ could be amplified till
$3{\times}10^{3}G$ in solar corona , this data allows to obtain a
magnetic Reynolds from expression (\ref{26}) as high as
$R{m}\approx{6{\times}10^{7}}$ which is much higher than the usual
Reynolds number used in numerical simulations \cite{14}. Note that
from this value
$\frac{B_{\phi}}{B_{\theta}}\approx{6{\times}10^{3}}$ and formula
(\ref{28}) one obtains a very small value for stretching factor of
$K^{2}\approx{1.6}$ as a very small deviation of the stretching
which is however able to yield a strong steady dynamo action. This
seems to be distinct from the result obtained by Livermore, Hughes
and Tobias where a best stretching seems not produce a efficient
dynamo action. But as shall be discussed next , this is only due to
a high Reynolds number considered here. Comparison between equation
(\ref{31}) and Ricca's equation
\begin{equation}
\frac{{B^{2}}_{s}}{{B^{2}}_{\theta}}\ge{\frac{K^{2}r}{Tw}}
\label{37}
\end{equation}
shows that they are very similar knowing that Tw represents the
twist of the magnetic flux tube. Of course a nonuniform stretch is
also present here. As was mention by Livermore et al , one of the
reasons that it is useful to use spherical coordinates is that, most
of the astrophysical and planetary objects are of spherical
morphology. Note that also in the case of flux tube non-uniform
stretching the Riemannian stretch of the line elements of twisted
flux tube and the thin one $K^{2}:=1$
\begin{equation}
d{s_{0}}^{2}= dr^{2}+r^{2}d{{\theta}_{R}}^{2}+ds^{2} \label{38}
\end{equation}
is given by
\begin{equation}
\frac{d{l}^{2}}{d{s^{2}}_{0}}=
1+(K^{2}-1)\frac{ds^{2}}{d{s_{0}}^{2}} \label{39}
\end{equation}
This expression shows that the stretch of the twisted flux tube
depends on the stretch of magnetic flux tube axis length ds. Besides
by writting this expression as
\begin{equation}
\frac{d{l}^{2}}{d{s^{2}}_{0}}=
1+(K^{2})\frac{ds^{2}}{d{s_{0}}^{2}}\approx{1+(r^{2}{\kappa}^{2}cos{\theta})\frac{ds^{2}}{d{s_{0}}^{2}}}
\label{40}
\end{equation}
which shows that the stretch also depends upon the Frenet curvature
and in the helical solar tube, where the Frenet curvature and
torsion coincides, this stretch depends also on the torsion of the
flux tube. Actually the non-uniform carachter of the flux tube was
actually discussed by Fisher et al who argued that: " Most likely
stretching mechanism is radial differential rotation, in the
overshooting layer, implying that flux tube does not lie at a single
depth and its properties are therefore uniform". This is exactly
what happens in the Riemannian model of solar dynamo presented here.
\newpage
\section{Conclusions} A particular solution of
self-induction and MHD equations, is found representing a steady
dynamo on flat uniformly stretched Riemannian manifold following
earlier Riemannian models of dynamos and magnetic flux tubes in
other systems of coordinates. It is shown that the Riemann space is
flat when the solar dynamo is unstretched, in close analogy with the
elastic stars in Einstein's general relativity. These ideas can help
us to build pseudo-Riemannian stretched relativistic models in
analogy with the 3D counterpart investigated here. The absence of
the radial component $B_{r}$ stems from the fact that in Ricca's
work \cite{3} it is also absent and since the present paper assumes
a analogous model for the solar dynamo in spherical coordinates this
assumption seems to be justifiable. Besides the assumption of
vanishing radial magnetic field component simplifies a great deal of
computation avoiding that one has to use numerical computation and
that a analytical solution can be obtained here. One important
feature to test the model discussed here from the solar physics
point of view is to include ohmic losses and diffusion as done by
Livermore et al \cite{9}. Though they found that stretching in
dynamo flows is not always a factor of enhancement and may even turn
the dynamo inefficient, they basically obtained their result for low
Reynolds numbers between $20<Rm<O(10^{2})$ while here one deals with
magnetic Reynolds numbers of the order of $Rm=O(10^{7})$. High Rm
numbers are actually related to the fast dynamos widely investigated
in the literature. In a future publication we shall address this and
other issues connected with spherical dynamos Riemannian manifolds
in detail. The solar cycle dynamos \cite{15} which stretches the
poloidal magnetic fields to toroidal magnetic ones by differential
rotation are therefore compatible with the spherical Riemann-flat
model discussed here.
\section{Acknowledgements}
I appreciate financial  supports from UERJ and CNPq.

\end{document}